\documentclass{aastex63}

\usepackage{url}
\usepackage{tabularx}
\usepackage{amsmath}
\usepackage{rotating}
\usepackage{multirow}
\usepackage{xcolor}
\shorttitle{Prediction of Solar Proton Events}
\shortauthors{Sadykov et al.}

\begin{document}

\title{Prediction of Solar Proton Events with Machine Learning: Comparison with Operational Forecasts and ``All-Clear'' Perspectives}
\email{vsadykov@gsu.edu}

\author[0000-0002-4001-1295]{Viacheslav M. Sadykov}
\affiliation{Physics \& Astronomy Department, Georgia State University, Atlanta, GA 30303, USA}

\author{Alexander G. Kosovichev}
\affiliation{Physics Department, New Jersey Institute of Technology, Newark, NJ 07102, USA}
\affiliation{NASA Ames Research Center, Moffett Field, CA 94035, USA}

\author{Irina N. Kitiashvili}
\affiliation{NASA Ames Research Center, Moffett Field, CA 94035, USA}

\author{Vincent Oria}
\affiliation{Computer Science Department, New Jersey Institute of Technology, Newark, NJ 07102, USA}

\author[0000-0003-2846-2453]{Gelu M. Nita}
\affiliation{Physics Department, New Jersey Institute of Technology, Newark, NJ 07102, USA}

\author{Egor Illarionov}
\affiliation{Department of Mechanics and Mathematics, Moscow State University, Moscow, 119991, Russia}
\affiliation{Moscow Center of Fundamental and Applied Mathematics, Moscow, 119234, Russia}

\author{Patrick M. O'Keefe}
\affiliation{Computer Science Department, New Jersey Institute of Technology, Newark, NJ 07102, USA}

\author{Yucheng Jiang}
\affiliation{Computer Science Department, New Jersey Institute of Technology, Newark, NJ 07102, USA}

\author{Sheldon H. Fereira}
\affiliation{Physics Department, New Jersey Institute of Technology, Newark, NJ 07102, USA}

\author{Aatiya Ali}
\affiliation{Physics \& Astronomy Department, Georgia State University, Atlanta, GA 30303, USA}

\begin{abstract}
	Solar Energetic Particle events (SEPs) are among the most dangerous transient phenomena of solar activity. As hazardous radiation, SEPs may affect the health of astronauts in outer space and adversely impact current and future space exploration. In this paper, we consider the problem of daily prediction of Solar Proton Events (SPEs) based on the characteristics of the magnetic fields in solar Active Regions (ARs), preceding soft X-ray and proton fluxes, and statistics of solar radio bursts. The machine learning (ML) algorithm uses an artificial neural network of custom architecture designed for whole-Sun input. The predictions of the ML model are compared with the SWPC NOAA operational forecasts of SPEs. Our preliminary results indicate that 1) for the AR-based predictions, it is necessary to take into account ARs at the western limb and on the far side of the Sun; 2) characteristics of the preceding proton flux represent the most valuable input for prediction; 3) daily median characteristics of ARs and the counts of type II, III, and IV radio bursts may be excluded from the forecast without performance loss; and 4) ML-based forecasts outperform SWPC NOAA forecasts in situations in which missing SPE events is very undesirable. The introduced approach indicates the possibility of developing robust ``all-clear'' SPE forecasts by employing machine learning methods.
\end{abstract}


\section{Introduction}
\label{Section:introduction}

    Solar Energetic Particle events (SEPs) are sharp increases in the particle flux from the Sun as observed in the near-earth environment. SEPs represent one of the clearest manifestations of transient solar activity and are one of the drivers of Space Weather disturbances. Accelerated directly at the sites of solar flares or at the shock wave fronts formed by Coronal Mass Ejections (CMEs), SEPs propagate through the Heliosphere and interact with the space environment of the Earth \citep{Reames2021}. There are various consequences of such interactions, including impacts on astronauts' health \citep{Martens2018}, increase in the radiation levels at aviation altitudes during the strongest events \citep{Kataoka2018} and corresponding economic impacts \citep{Saito2021}, damage to satellites, and potential danger for space exploration. Consequently, monitoring and prediction of SEPs are high priorities for the operational and research communities.
    
    Solar Proton Events (SPEs) represent a major subclass of SEP events. A classical definition of the least-strength S1 events (on the scale of the Space Weather Prediction Center at the National Oceanic and Atmospheric Administration \citep[][]{NOAA2021_scales}) is an enhancement of the flux of $>$\,10\,MeV protons above the 10\,pfu (particle flux unit) threshold. Prediction of such SPEs is a challenging problem from many perspectives. During solar cycle 24, one or more such enhancements were observed less often than 1 day in 30. This ratio (defined here as a class-imbalance) becomes even stronger for higher fluxes or energy thresholds. Difficulties and strategies for forecasting space weather events for highly-imbalanced data sets were pointed out by \citet{Ahmadzadeh2021ApJS-howtotrain}, and a need for balanced data sets for space weather prediction was emphasized by \citep{Martens2018b}. In addition, SPEs may be significantly delayed (up to tens of hours) with respect to the onset time of the preceding solar flare. Therefore, the forecasting time window and data handling are challenging because of the different timescales for processes associated with SPEs. Also, it is important to note that the origin of SPEs at the solar surface (or the exact location of the host Active Region, AR) sometimes cannot be determined, especially if the host AR is very close to or behind the western limb of the Sun. In this situation, the characteristics of the magnetic field structure and complexity in the host AR cannot be defined. The last two points make it challenging to connect a particular SPE to its host AR.
    
    With the increasing amount and quality of available data and the development and accessibility of machine learning (ML) algorithms, there has been a growing number of attempts to apply ML to predict solar transient events. For example, \citet{Huang2012} utilized the ensemble forecasting approach for SPEs based on preceding flares and CMEs and found better performance if these data are combined to make forecasts. \citet{Kim2018} developed an algorithm to predict SPEs using the solar radio flux by employing statistical analysis, neural networks, and genetic algorithms. \citet{Jeong2014} developed an algorithm of SPE forecasting with NOAA scales based on Multi-Layer Perceptrons (MLPs) and using GOES solar soft X-ray (SXR) characteristics of flares from 1976 to 2011. \citet{Bain2019} pointed that ML models built on the data available in real-time could outperform operational forecasts. In addition, there is growing interest in the development of forecasting frameworks \citep{Engell2017} and ML-ready data-set preparation \citep{Martens2018b}.
    
    The current operational daily forecasting of $>$\,10\,MeV $>$\,10\,pfu SPEs relies on a statistics-based approach, plus a human forecaster to correct the predictions \citep{Balch1999,Balch2008,Bain2020-SEPNOAA}. This approach has been validated over several solar cycles. Despite advances in ML-based predictions, comparisons with these operational SWPC NOAA forecasts or application to the same temporal and spatial scales \citep{Zhong2019,Jeong2014,Bain2019} are rarely made.
    
    In addition to the daily forecasting of $>$\,10\,MeV $>$\,10\,pfu proton events, there are approaches targeted on shorter timescales for the prediction \citep{Laurenza2018,Papaioannou2018}, higher proton energy thresholds \citep{Nunez2015,Malandraki2018}, and various quantities related to the SPEs \citep[fluence, peak flux, etc.,][]{AminalragiaGiamini2020,Lovelace2018}. One of the recent focuses of SPE predictions is the ``all-clear'' approach, where a missed event is considered a significant failure of the algorithm as compared to a false alarm event. This approach is of special importance for mitigating risks from radiation exposure.
    
    In this work, we present a novel approach for developing daily prediction capabilities of Solar Proton Events (SPEs) based on entire Sun observations, applying an artificial neural network of custom architecture, and comparing the prediction results with the SPE forecasts by SWPC NOAA. We also extend this approach to the ``all-clear'' prediction problem for SPEs. Section~\ref{Section:datapreparation} describes the data preparation, derivation of characteristics, and data set construction for the prediction. Section~\ref{Section:mlsetup} presents the architecture of the artificial neural network for SPE prediction and the training strategy. Section~\ref{Section:results} summarizes the main results of this work and is followed by a discussion in Section~\ref{Section:discussion}. The conclusion is presented in Section~\ref{Section:conclusion}.

\section{Data Preparation}
\label{Section:datapreparation}

    The data processing and machine learning setups described in Sections~\ref{Section:datapreparation} and~\ref{Section:mlsetup} are publicly available at the \citet{Sadykov2021-GoogleDriveLink}.

    \subsection{Daily SWPC NOAA Proton Event Probabilities}
    
        Operational space weather forecasts of solar transient events are currently prepared jointly by the Space Weather Prediction Center (SWPC) of the National Oceanic and Atmospheric Administration (NOAA) and the 557th Weather Wing of the United States Air Force (USAF). The forecasts are released daily, at 22:00\,UT, for the three following days (starting from 00:00\,UT). The forecasts are available via ftp (\url{ftp://ftp.swpc.noaa.gov/pub/warehouse/}) in the RSGA folder for each year. The forecast of solar proton events is presented in section ``III'' of that document in terms of probabilities (ranging from 1 to 99) of the $>$\,10\,MeV proton flux to exceed 10\,pfu. The forecasts rely on statistical models \citep{Balch1999,Balch2008} augmented by the “forecaster-in-the-loop” technique (the calculated probabilities are corrected by forecasters based on the forecasters' experience). This technique was proven to have a higher performance than non-corrected probabilities for the SXR solar flares and in terms of Brier Skill Score \citep[BSS,][]{Crown2012}. For the current study, we utilize only the next-day probabilities of the proton events. The data used were collected for each day from June 2010~--- December 2019. The probabilities for the days when the reports were not issued were obtained by interpolation from the neighboring days. An extensive study of the SWPC NOAA SPE probabilities was recently performed by \citet{Bain2020-SEPNOAA}; we direct the readers to this study for more details about the operational forecasts.
        
        One of the goals of this study is to compare the performance of ML-based prediction of SPEs with the probabilistic operational forecasts. Because the next-day SWPC NOAA probabilities are issued at 22:00\,UT and are valid starting from 00:00\,UT, we will adopt the following convention in this study: all characteristics described below will be computed for each considered day from the data acquired between 22:00\,UT of the previous day and 22:00\,UT of the considered day, i.e., they can be fully defined at 22:00\,UT of the considered day. However, the prediction will be made for the presence of $>$10\,MeV $>$10\,pfu events during the next day starting from 00:00\,UT, i.e., the forecast will have a 2-hour latency time window. This setting will allow us to match the timing of SWPC NOAA operational forecasts.

    \subsection{Records of Radio Bursts}

        Radio bursts of types II and IV are prominent indicators of the formation of shocks in the Heliosphere \citep{Zimovets2015}, and, together with type III radio bursts, are valuable characteristics for SEP forecasts \citep{Winter2015,Miteva2017,Ameri2019,Bain2019,Kalaivani2020}. Therefore, information about such radio bursts is essential to consider for forecast development. In this work, we utilize a simplistic approach by counting the number of type II, III, and IV radio bursts that occurred during each day. The daily counts of the radio bursts will be utilized as characteristics to predict SPEs for the following day. The records of the radio bursts are publicly available at SWPC NOAA via FTP (\url{ftp://ftp.swpc.noaa.gov/pub/warehouse/}).

    \subsection{Daily Median SHARP Characteristics of Active Regions}
        
        The characteristics of the vector magnetic fields in ARs (such as the total unsigned magnetic helicity, total unsigned vertical electric currents, etc.) are often used to predict solar flares and CMEs \citep[e.g.,][]{Bobra2015,Bobra2016,Campi2019}. Space Weather HMI Active Region Patches \citep[SHARPs,][]{Bobra2014} are, probably, the most widely-used set of vector magnetic field characteristics calculated for each recognized magnetic field patch with a 12\,min cadence. In this study, the ``Active Regions'' (ARs) term will correspond to any magnetic field patches recorded in the SHARP database. We compute the daily median values of the Cylindrical Equal Area (CEA) projected SHARP characteristics for each patch and use these for the prediction of SPEs. There are 30 characteristics for each magnetic patch used in this study, including the patch center coordinates and data quality parameters.
       
        The SHARP characteristics are reliable only if the AR is located within $\approx$68\,deg from the solar disc center \citep{Bobra2015}, due to projection effects and the loss of the spatial resolution closer to the limb. This represents a problem for SPE prediction. In Parker's solar wind model \citep{Parker1958}, the solar longitudes magnetically connected to the Earth are located close to the western solar limb. Statistical studies of the distribution of SPE origins \citep{Cliver2020} also demonstrate an asymmetry towards the western limb, with the average SPE origin range of east 20 – west 100 longitudes. As a result, one has to consider patches in close proximity to the western limb and beyond to issue a whole-Sun prediction of SPEs.
        
        To address this issue, we mimic the existence of the magnetic field characteristics close to and behind the western solar limb by assuming that the AR has the same magnetic field characteristics as for the previous day. We also assume that the AR obeys the Carrington rotation rate ($\sim$25.4~days) and exists on the far side of the Sun for 11 more days while preserving magnetic field characteristics. We apply this strategy for every magnetic field patch recognized in SHARP.
        
        The paper aims to utilize the characteristics of the magnetic field of patches (ARs) over the whole Sun to predict the SPEs. The number of individual ARs (or magnetic patches) on the Sun varies from day to day. It is known that large and more complex ARs tend to produce more flares and SPEs  \citep[e.g.,][]{Falconer2012,Sadykov2017}. Therefore, for each day, we select 10 SHARPs with the largest unsigned magnetic field fluxes (including those extrapolated to and behind the western limb) and use their characteristics for the whole-Sun prediction. In case there are less than 10 patches recorded for the current day, we introduce zero entries for the absent patches.

    \subsection{Daily Characteristics of GOES Proton and Soft X-ray Flux Measurements}

        The recent history of transient solar activity (e.g., the number of prior transient events within one or several days) may serve as a useful characteristic for predicting future solar transient events \citep{Falconer2012,Nishizuka2017}. In our forecast of SPEs, we include information about prior activity in the form of the daily characteristics of soft X-ray (SXR) 0.5-4\,$\AA$ and 1-8\,$\AA$ fluxes and $>$\,10\,MeV proton fluxes, both observed by the GOES series. We have utilized the 1-minute averaged SXR fluxes and 5-minute averaged proton fluxes, available at the NOAA National Center for Environmental Information (NOAA NCEI, \url{https://satdat.ngdc.noaa.gov/sem/goes/data/avg/}). The proton flux observations are done by GOES in eastward and westward directions. In this study, we utilize the average flux from the two detectors. We also discuss in Section~\ref{Section:discussion} how using the flux from only one of two detectors affects the results.
        
        For each day from June 2010~--- December 2019, we compute the mean, median, minimum, and maximum fluxes of 0.5-4\,$\AA$ and 1-8\,$\AA$ SXR radiation and $>$\,10\,MeV protons, as well as their daily standard deviations. We also record the lowest available proton flux for the day (i.e., at 22:00\,UT).

    \subsection{Normalization of Characteristics, Labeling, Oversampling, and Train-Test Separation}

        Extraction of characteristics yields, for each day, 300 characteristics corresponding to the magnetic fields of ARs (each of 10 included ARs/patches has 30 characteristics), 10 SXR characteristics (5 for each SXR channel), 6 characteristics of the $>$10\,MeV proton flux, and counts of type II, III, and IV radio bursts. Each characteristic is normalized to the zero-one range, the similar AR characteristics (for example, the unsigned current helicities for each of 10 ARs) are normalized following the same scale. These characteristics are organized into 319-component vectors, one for each day. Each vector of characteristics is assigned with the label, 1 or 0, depending on whether the flux of $>$\,10\,MeV protons (averaged from two GOES detectors) hits the 10\,pfu threshold sometime during the next day or not. As a result of the labeling, there were 101 days during the June 2010~--- December 2019 time period labeled as days with SPEs and 3400 days labeled as having no SPEs. That leaves us with a class-imbalance ratio of about 1 to 34.
        
        Next, we separate the labeled vectors of characteristics into the train and test data subsets. The separation strategy shown in Table~\ref{table1}. As one can see, there are approximately twice as many samples in the train data set than in the test data set, and the class-imbalance ratios in both data sets are also kept approximately equal to 1 to 34. Because the separation is done by keeping long chunks in the same data sets, we avoid the temporal coherence problem \citep{Ahmadzadeh2021ApJS-howtotrain}.
        
        Because the data set suffers from class-imbalance, we adopt a strategy to mitigate this \citep{Ahmadzadeh2021ApJS-howtotrain}. In this work, we oversample the positive cases (days with SPEs) in the dataset to balance the number of negative cases. In essence, this means that we introduce 34 times more positive samples into the data set. In this case, both the train and the test data sets become balanced, which eases the ML training by allowing us to use any common loss function for training.

\section{Machine Learning Setup}
\label{Section:mlsetup}

    There are several challenges in SPE prediction, including the unknown origin on the solar surface for some events, and difficulties in separating the proton flux enhancement from variations of the proton flux background. These problems could be solved if we consider the whole Sun for building a prediction. There is also the third issue. The number of days with SPEs (66 days in the train data set, see Table~\ref{table1}) is small in comparison to the number of characteristics (319 for each day). In such situations, overfitting (i.e., memorization of the train data set instead of its generalization) is a common problem. To avoid overfitting at the level of the neural network architecture, one can bring some physical guidance into the prediction algorithm and reduce the number of free parameters to tune during the training. Here, we present some guidance on how the neural network architecture should look. We note, first, that the data set contains the characteristics of 10 ARs with the largest unsigned magnetic fluxes and the pre-processing should be the same for each of the 10 ARs used for prediction. Second, the prediction of SPEs and other transient events is typically performed for individual ARs using highly correlated characteristics of the magnetic field. Both of these points help us determine the neural network architecture to be used for the whole-Sun problem.
    
\begin{figure}[ht]
	\centering
	\includegraphics[width=1.0\linewidth]{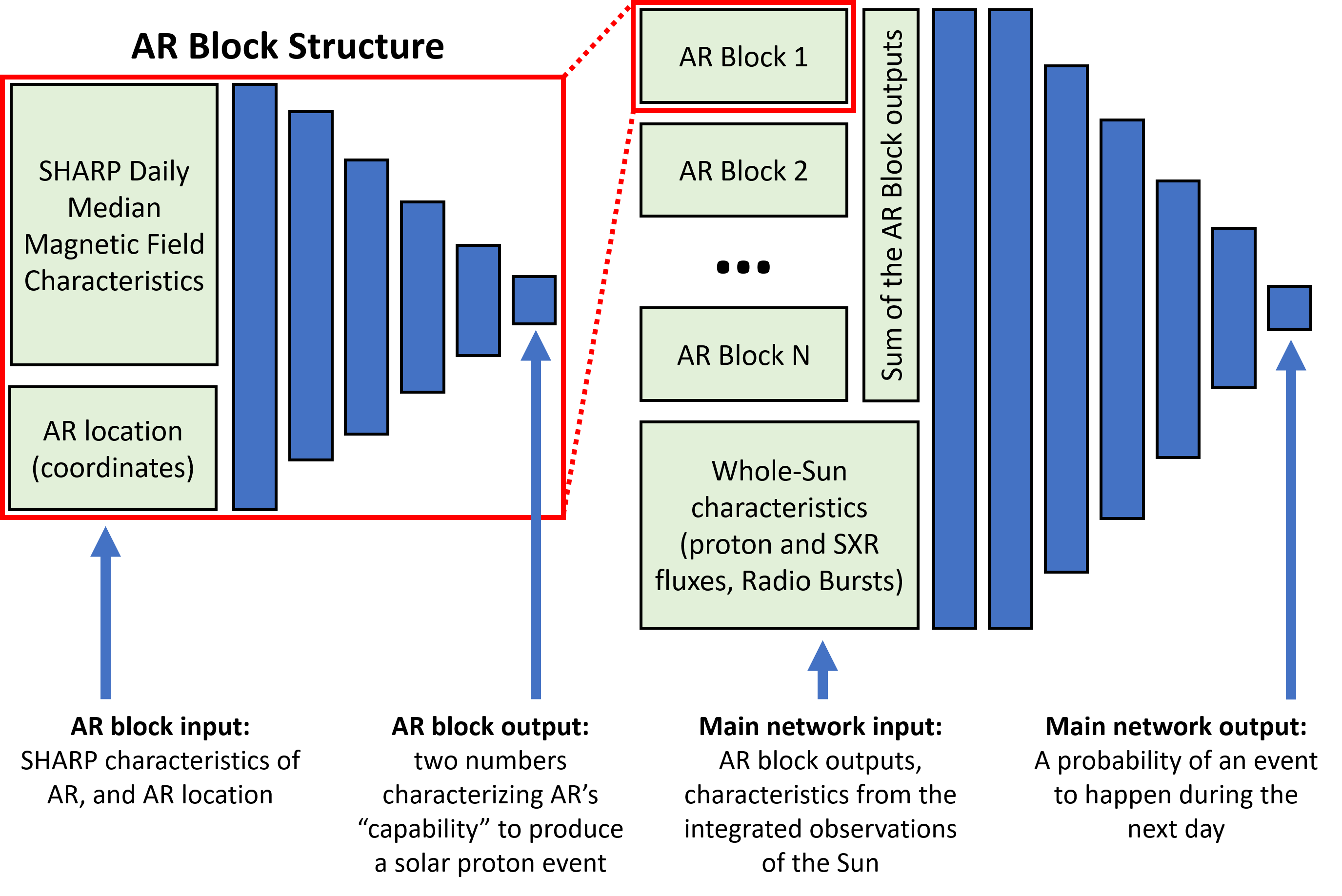}
	\caption{A schematic representation of a neural network architecture for daily whole-Sun prediction of SPEs. The number of AR Blocks is equal to 10 for this work. The neuron connections in AR Blocks share the weights and the biases.}
    \label{figure1}
\end{figure}

\begin{figure}[ht]
	\centering
	\includegraphics[width=1.0\linewidth]{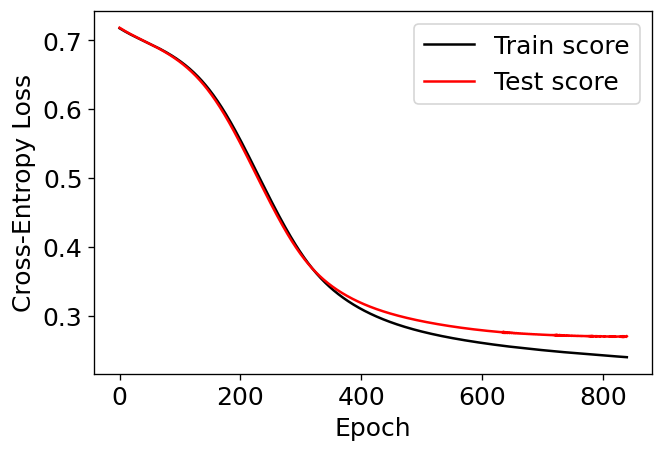}
	\caption{An example of the cross-entropy loss function decay obtained for the train and test subsets during the neural network training procedure.}
    \label{figure:training}
\end{figure}

    Figure~\ref{figure1} illustrates the architecture of the neural network used in this work. As one can see, the characteristics for each AR are initially pre-processed in the form of ``AR blocks''~--- sets of fully connected layers that reduce the initially entered 30 characteristics of each AR to 2 characteristics summarizing the AR capability to produce an SPE. The numbers of neurons in the subsequent layers of the AR blocks are set at 30 - 15 - 8 - 4 - 2. All AR blocks share the same weights and biases (free parameters tuned during the neural network training). The outputs of the blocks are then summed up and entered in the main part of the network together with other daily characteristics (SXR and $>$\,10\,MeV proton flux characteristics, and the counts of type II, III, and IV radio bursts). The numbers of neurons in subsequent layers of the main part are set at 21 - 15 - 10 - 5 - 2. The output of the network consists of two numbers. Depending on which of these numbers is larger, the algorithm issues a prediction of the occurrence/non-occurrence of an SPE event during the next day. Rectified Linear Unit (ReLU) activation functions are used everywhere in the network, and the two outputs of the network are normalized to the probabilities using the Soft-Max function. The loss function for network training is the cross-entropy loss defined as:
    \begin{gather}
	\label{eq:crossentropy}
	    \mathcal{L} = -\dfrac{1}{N}\sum_{i=1}^{N}y_{i}\cdot{}log(p_{i}) + (1-y_{i})log(1-p_{i}).
	\end{gather}
    Here $N$ is the total number of samples in the data set, $y_{i}$ is the label of the $i$-th feature vector (1 or 0), and $p_{i}$ is the predicted probability of the $i$-th feature vector to have a label 1. The ``adam'' optimizer \citep{Kingma2015} with a learning rate of 0.00025 was used to progress the training. The batch size was equal to the entire training data set (i.e., the data set was not subdivided into the ``mini-batches''). To avoid overfitting while training the network, we employ the early stopping criterion. After each iteration (epoch) of the training process, we evaluate the performance of the network on the test data set in terms of the loss function $\mathcal{L}_{test}$. We stop the learning process when the mean value of $\mathcal{L}_{test}$ for ten consecutive epochs is larger than or equal to the mean value of $\mathcal{L}_{test}$ for the previous five consecutive epochs. The same training procedure is repeated five times to check if weight initialization and data set shuffling affect the training and the resulting prediction. An example of the training process for the neural network trained on all characteristics is presented in Figure~\ref{figure:training}.
    
    To test how the inclusion or exclusion of various characteristics affects the forecast, we substitute the excluded characteristics with a certain constant number (equal to 0.5) without modifying the network architecture. We also examine the network performance using various metrics for binary predictions. For binary predictions, we utilize the True Skill Statistics (TSS) and Heidke Skill Score (HSS) metrics of the 1st and 2nd kind \citep{Bobra2015}. The optimal threshold for the network output is selected for each particular binary metric separately. In addition, we look at the Receiver Operating Characteristic (ROC) curve and calculate the area under this curve. The ROC curve illustrates how the false-positive rate (the fraction of incorrectly-predicted days without SPEs) and true-positive rate (the fraction of correctly-predicted days with SPEs) change with the variation of the threshold for the network output. Because we are also interested in all-clear forecasts, we introduce a new metric, Weighted True Skill Statistics (WTSS), defined as:
    \begin{gather}
	\label{eq:WTSS}
    	WTSS(\alpha{}) = 1 - \dfrac{2}{\alpha{}+1}\left(\alpha{}\dfrac{FN}{P}+\dfrac{FP}{N}\right), \\
    	WTSS(1) = TSS = 1 - \dfrac{FN}{P} - \dfrac{FP}{N}.
	\end{gather}
	Here FN is the number of SPEs predicted as no-SPEs, FP is the number of no-SPEs predicted as SPEs, P is the total number of SPEs, N is the total number of no-SPEs, $\alpha{}$ is the parameter that indicates how strongly the missed event rate (FN/P) is penalized with respect to the false alarm rate (FP/N). The WTSS metric preserves some of the properties of the TSS metrics. First, for each particular $\alpha{}$, it is not sensitive to the class-imbalance ratio and ranges from -1 (totally wrong forecast) to 1 (entirely correct forecast). WTSS = 0 corresponds to the random choice forecast. We analyze how this score behaves for different $\alpha$ values. The large values of $\alpha$ are of special interest for the ``all-clear'' forecasts as they indicate a significantly lower tolerance to the missed event rate than the false alarm rate.

\section{Results}
\label{Section:results}

    \subsection{Comparison of Neural Network Results with SWPC NOAA}
    
        A comparison of the prediction results obtained with the ML algorithm (an artificial neural network) with the daily operational forecasts by SWPC NOAA and persistence forecast (the prediction for the next day being the same as the observation for the current day) in terms of various metrics is summarized in Table~\ref{table:comparisonSWPC}. Based on these results, the ``winner'' (the forecast receiving the highest score) depends on the selected metrics. For example, the SWPC NOAA forecast outperforms the ML-based forecast in terms of the HSS1 and HSS2 metrics (i.e. has higher scores for these metrics). However, the situation is the opposite for almost all other metrics (including the TSS score), where the ML-based forecast performs better than the SWPC forecast. The persistence model performs worse for both the operational forecast and ML-based model prediction.
        
\begin{figure}[ht]
	\centering
	\includegraphics[width=1.0\linewidth]{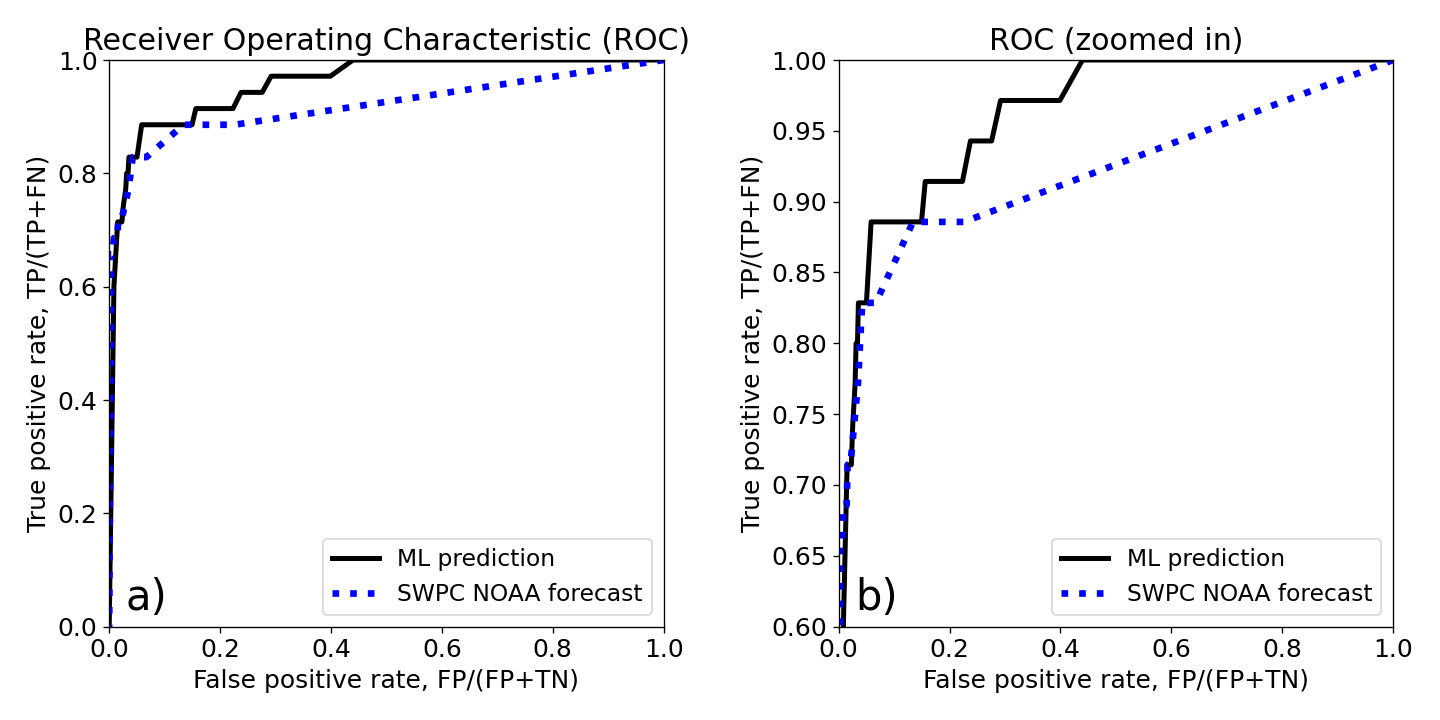}
	\caption{(a) Receiver Operating Characteristic (ROC) curves for the machine learning-based forecast using all characteristics (black solid line) and for the SWPC NOAA forecast (blue dotted line). Panel (b) presents the zoomed version of the same ROC curves.}
    \label{figure:ROCgen}
\end{figure}

        Figure~\ref{figure:ROCgen} illustrates the ROC curves for the ML-based prediction and SWPC NOAA operational forecast. These ROC curves look very similar to the ones presented by \citet{Huang2012} in Fig. 4. One can see that the primary difference between the two ROC curves is in the region of high true-positive rates (and, correspondingly, low false-negative rates). There, the ML-based forecast outperforms the SWPC NOAA forecast. Because the top of the ROC curve corresponds to the region when the number of false-negative predictions is very low (i.e., when very few SPEs are missed), this region can be considered as a region of the ``all-clear'' forecasts. Correspondingly, the ML-based predictions outperform the SWPC NOAA forecasts in terms of the potential to issue ``all-clear'' predictions. This is also clearly visible from the analysis of WTSS scores in Table~\ref{table:comparisonSWPC} and is discussed later in the text.
        
\begin{figure}[ht]
	\centering
	\includegraphics[width=1.0\linewidth]{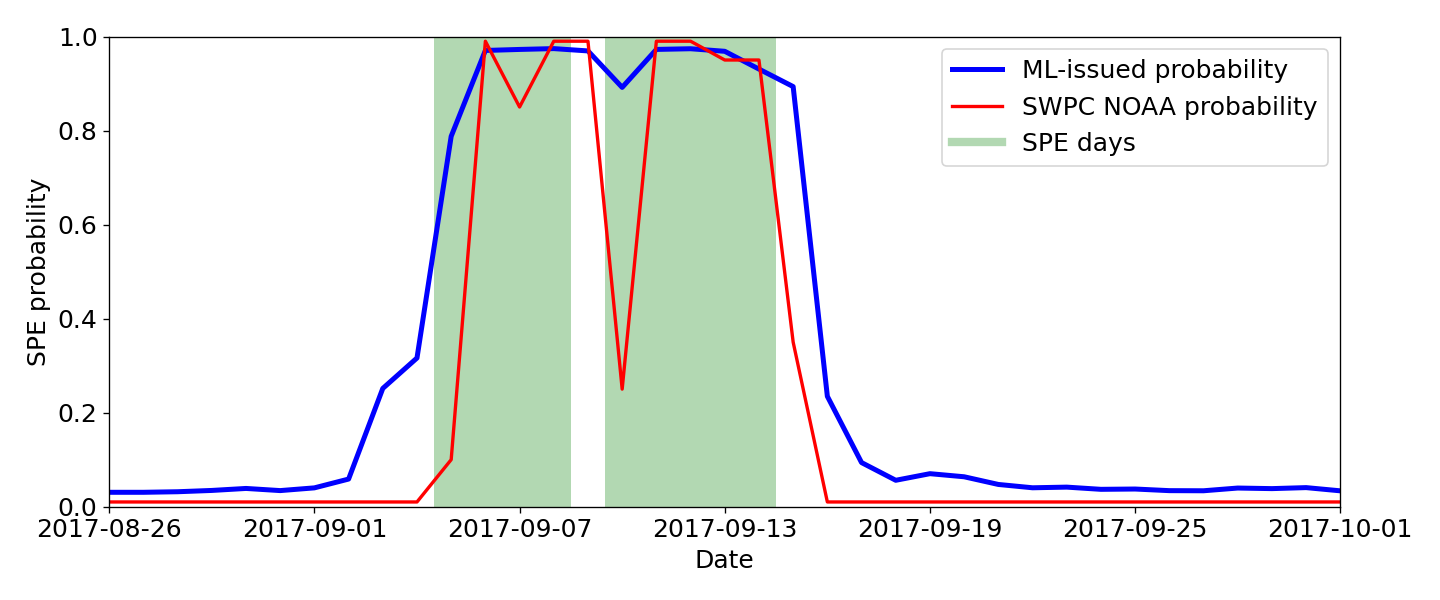}
	\caption{Probability of the SPEs rolled over time for August 26 - October 1, 2017 time period. The red line corresponds to the SPE probability issued by SWPC NOAA for the current day, the blue line corresponds to the machine learning-issued probability for the current day, the green rectangles mark actual times of SPE occurrence.}
    \label{figure:timeroll}
\end{figure}

        Another way of comparing the forecasts is to unroll the probabilities over time and compare them for particular events \citep{Benvenuto2020}. Figure~\ref{figure:timeroll} illustrates how the SWPC NOAA forecast and ML-based prediction performed during the September 2017 events. The soft-max function was used to normalize the output of the neural network (two channels) into ``probabilities''. Although these numbers may not necessarily have a meaning of probability in the statistical sense, they are still enough to analyze how the network reacts to the upcoming event. One can see two major SPEs during that month, marked by green rectangles in the plot. It is clear that the SWPC NOAA forecast reacts to both events with one day delay, while the ML-issued forecast captures the beginning of the first event well, and the corresponding probability does not drop strongly at the beginning of the second event.
        
\begin{figure}[ht]
	\centering
	\includegraphics[width=1.0\linewidth]{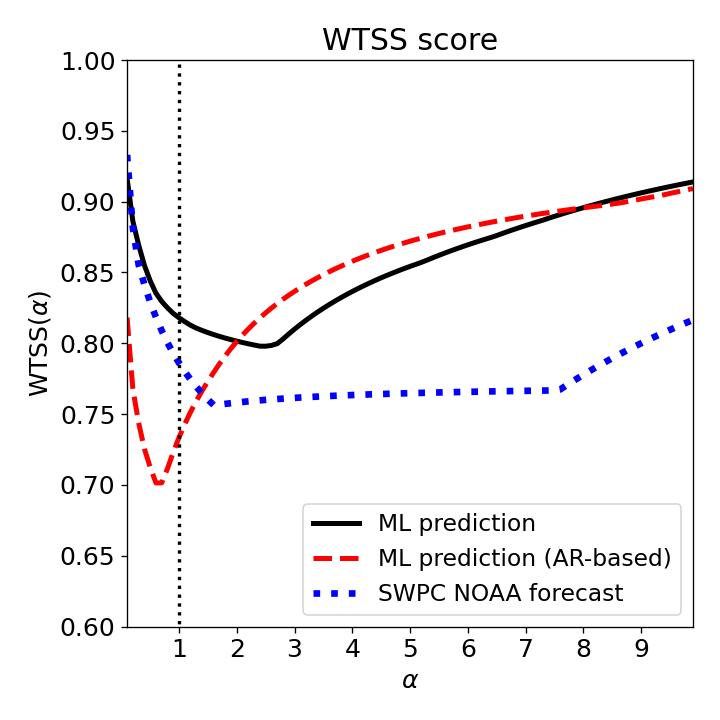}
	\caption{The Weighted True Skill Statistics (WTSS) as a function of a weighting parameter ($\alpha$) for ML-based predictions using all characteristics (black solid line), using SHARP characteristics of ARs only (red dashed line), and calculated for the SWPC NOAA probabilities (blue dotted line). The vertical black dotted line corresponds to $\alpha{}=1$ case, when $WTSS(1)=TSS$.}
	\label{figure:WTSS}
\end{figure}

        In the previous chapter, we have introduced the Weighted True Skill Statistics (WTSS) metrics. The parameter $\alpha$ determines the ratio of penalties for the missed event rate and the false alarm rate. Figure~\ref{figure:WTSS} illustrates the WTSS score as a function of $\alpha$  for the ML-based prediction and the SWPC NOAA forecasts. Any value of $\alpha{} > 1$ indicates the preference to have a higher false alarm rate with respect to the missed event rate. One can see in this figure that, while the ML prediction based on all characteristics and the SWPC NOAA forecast are mostly similar to each other for $\alpha{} < 1$, the ML-based prediction starts to outperform the SWPC NOAA forecast for $\alpha{}\geq{}1$. This indicates that the ML-based forecasts behave better in the regime when the missed events are not desirable (i.e., in the ``all-clear'' regime).

    \subsection{Prediction Based on SHARP AR Characteristics}
    
        In addition to building an ML-based model on all available characteristics, we also investigate the model performance based on AR characteristics only (i.e., SHARP characteristics for this research). We consider four cases here: 1) learning based on the SHARP characteristics where the extension of ARs behind the limb is taken into account; 2) learning without such extension; 3) learning when the AR coordinates are not provided to the neural network; and 4) learning then the ARs are distinguished based on their unsigned magnetic flux (i.e., not summed up before entering the main network, see Section~\ref{Section:mlsetup} and Figure~\ref{figure1}). The results are summaries in rows 2-5 in Table~\ref{table3}.
        
\begin{figure}[ht]
	\centering
	\includegraphics[width=1.0\linewidth]{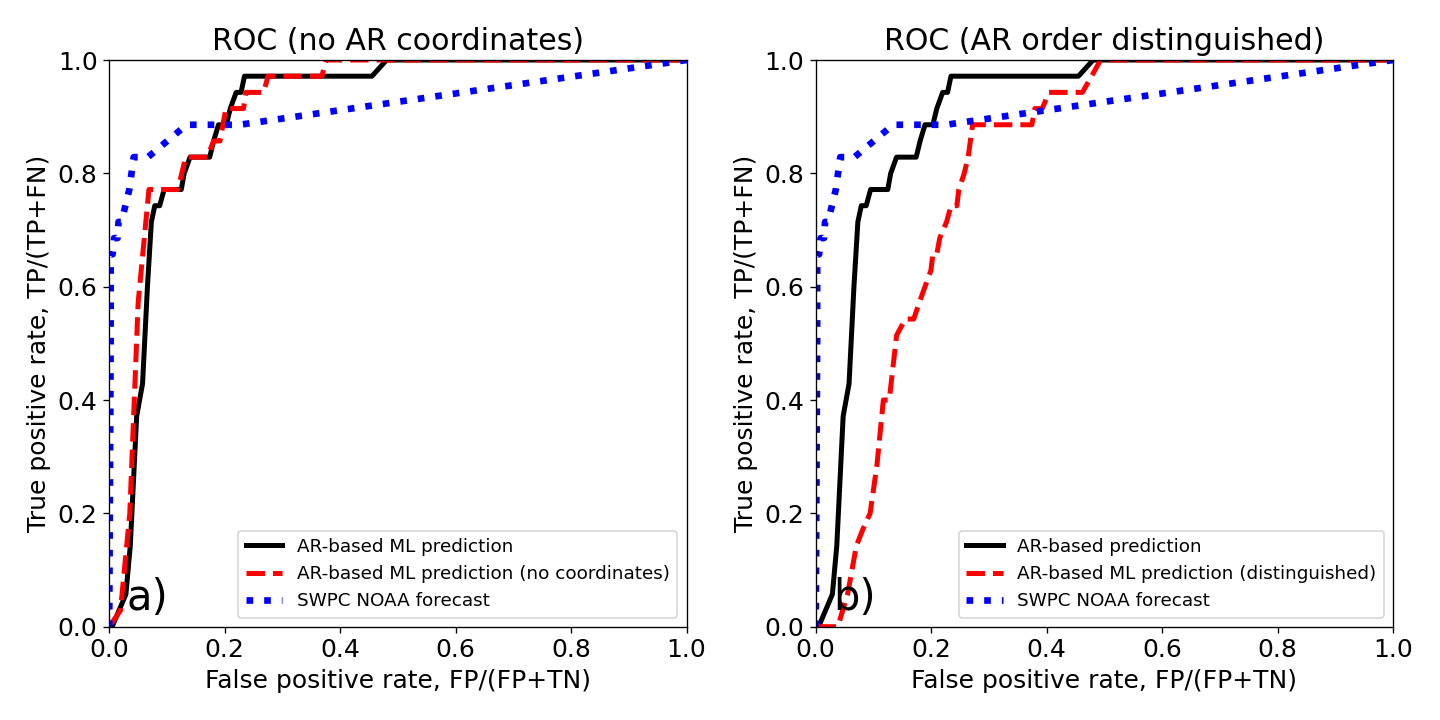}
	\caption{Comparison of the ROC curves for the ML predictions based on AR characteristics only (black solid line) and SWPC NOAA forecasts (blue dotted line). Additional red dashed lines correspond to the cases when no AR coordinates were introduced to the neural network (panel a), and when no AR extension to an behind the western limb is made (panel b).}
  	\label{figure:ROC-ARbased}
\end{figure}

        The AR-based predictions score is smaller compared to SWPC NOAA forecasts for any of the scores considered in Table~\ref{table3} except WTSS for $\alpha{}\geq{}2$. For example, the HSS1 score (indicates the performance of the prediction with respect to the fully negative no-SPE prediction) is zero for the AR-based predictions. Other scores also experience a decrease when only SHARP characteristics are considered. Such a decrease in performance is also evident from the ROC curves presented in Figure~\ref{figure:ROC-ARbased}. Another observation is that the extension of ARs to the western limb and beyond helps obtain a significantly better prediction compared to the case when ARs are not extended. This is evident from the scores in Table~\ref{table3} and the ROC curves in Figure~\ref{figure:ROC-ARbased}. The difference between the AR-based forecasts with and without the introduction of AR coordinates is negligible and, most probably, is related to the poor sampling of the positive events in the data and its small size. The last interesting point is, if the AR Block outputs are not summed before propagating to the main network, the performance of the forecast decreases in terms of all considered scores.
        
        Nevertheless, an interesting detail is that, although the AR-based forecasts do not perform well compared to SWPC NOAA for most of the considered metrics, they start to outperform the operational forecasts when approaching the ``all-clear'' regime. This is also evident in Figure~\ref{figure:WTSS}, where one can see that the AR-based forecast demonstrates higher $WTSS$ scores for $\alpha{}\gtrsim1.5$. The possible interpretation of such behavior is discussed in more detail in Section~\ref{Section:discussion}.

    \subsection{Influence of Feature Inclusion/Exclusion on the Forecast}
    
        In addition to considering AR-based predictions, we investigate how the exclusion of characteristics of various types affects the forecast. If the exclusion of the characters from the data set does not affect the predictor performance after its retraining, the information carried by this characteristic with respect to the predicted quantity can be inferred from other characteristics. On the other hand, if the predictor performance drops significantly, the characteristic cannot be replaced by other inputs and is especially valuable for the forecasts. First, we exclude the AR (SHARP) characteristics, the solar proton flux-related characteristics, the soft X-ray characteristics, and the counts of the radio bursts, each separately. The results of this experiment are summarized in Table~\ref{table3}. The exclusion of the proton flux characteristics seriously reduces the forecasting scores, and excluding other characteristics does not affect the score. On the other hand, predictions based solely on the proton flux characteristics have  slightly reduced performance. While there is no doubt that the prehistory of the proton flux is the most valuable characteristic, identifying the second most valuable one remains open.
        
        To answer this question, we considered each of the remaining groups of characteristics combined with the proton flux characteristics. The results are summarized in Table~\ref{table3} and visualized in Figure~\ref{figure:ROCvar}. The proton and SXR flux characteristics, grouped together, already reach and even outperform the prediction based on all characteristics. Predictions based on the proton flux and AR characteristics demonstrate a slightly weaker performance, and the results are even weaker for the inclusion of the counts of the radio bursts.

\section{Discussion}
\label{Section:discussion}
        
    One of the current goals is to highlight several new approaches to predict Solar Proton Events (SPE) and solar transient events in general. One such approach is a custom-built neural network used to construct whole-Sun-based predictions instead of considering each AR separately. In our opinion, there are several advantages to this approach. First, it naturally considers characteristics of the data integrated over the whole Sun and in-situ measurements, such as the SXR and proton fluxes. There is no need to link the events detected in such measurements to the host ARs. Therefore, there is no problem with the background for the events: any increase of the measured flux above a certain threshold may be considered as an event. This may be especially important for binary prediction of solar flares, where the estimate of the pre-flare background depends on the background subtraction method \citep{Ryan2012,Sadykov2019}.
        
    The second important advantage of the built architecture is the use of shared weights for ``AR Blocks'' and similar processing for ARs. This allows us to significantly reduce the number of free parameters in the network (weights and biases) and simplify the training. In fact, we have also considered a fully connected architecture with no weight sharing.  We found that a fully connected architecture is hard to train because of fast overfitting. Correspondingly, the shared-weight architecture introduced here was necessary to train the network successfully. The distinction between ARs within the network (which introduces more free parameters) makes the AR-based forecast worse (Table~\ref{table3}). 
        
    Our results showed that it is necessary to account for the ARs close to the western limb and on the far side for the AR-based predictions. Figure~\ref{figure:ROC-ARbased}\textit{b} and Table~\ref{table3} support this point. If no AR extension to and behind the western limb is applied, the forecast is worse in terms of the majority of the metrics, and the cross-entropy value is very close to the totally-random forecast value ($\sim$0.69). Because the solar longitude magnetically connected to the Earth is very close to the western limb ($\sim{}$78\,deg), these results are understandable. It was also pointed out by \citet{Cliver2020} that the SPE-producing ARs are located between 20 east - 100 west longitudes and tend to appear more in the western hemisphere. Therefore, a significant number of SPEs should come from the ARs whose characteristics are formally not ``trustable'' or even not known (i.e., the ARs are located farther than 68\,deg from the solar disc center). As one sees, even a simple model of the AR extension to and behind the western limb helps to improve the forecast. A more advanced model for the evolution of the AR characteristics can help build more robust forecasts.
        
\begin{figure}[ht]
	\centering
	\includegraphics[width=1.0\linewidth]{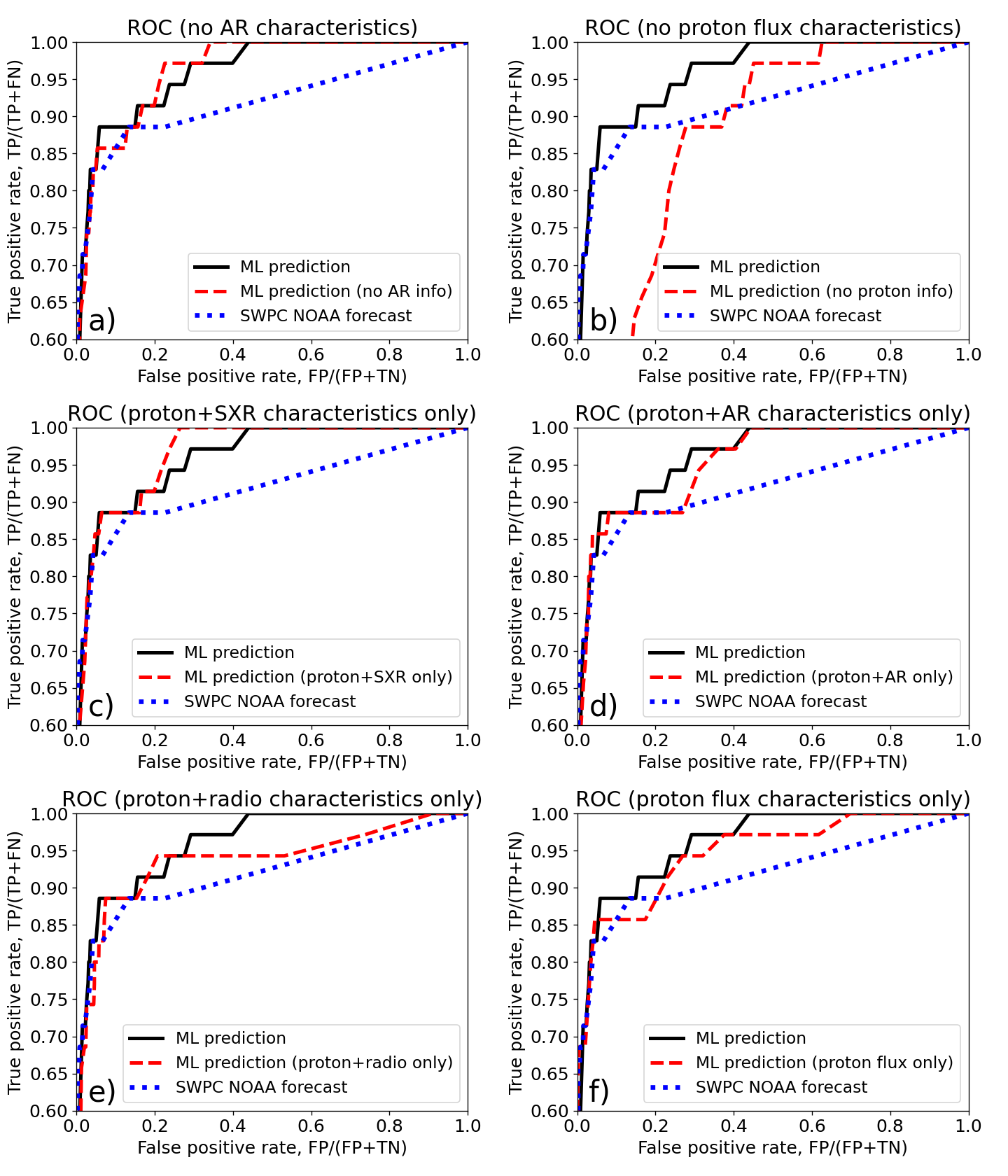}
	\caption{Comparison of the ROC curves for the ML predictions using all characteristics (black solid line) and SWPC NOAA forecasts (blue dotted line). Additional red dashed lines correspond to the cases where the network was trained with no AR characteristics (a), no proton flux characteristics (b), only on proton and soft X-ray flux characteristics (c), proton flux and AR characteristics (d), only on proton flux and radio burst characteristics (e), and solely on proton flux characteristics (f).}
  	\label{figure:ROCvar}
\end{figure}

    Figure~\ref{figure:ROC-ARbased}\textit{a} and Table~\ref{table3} also indicate that the exclusion of the AR coordinates does not make the forecast worse, which is a bit of surprise. Moreover, as evident in Table~\ref{table3} and Figure~\ref{figure:ROCvar}\textit{a}, the ML-based prediction does not worsen if the AR characteristics are completely excluded from the forecast. There are several possible explanations for such unexpected results. First, this may point to a low complexity of the implemented approach. In this study, we have worked only with the daily median values of SHARP characteristics but did not include their variations with time. A consideration of the evolution of SHARP characteristics and analysis of the corresponding multivariate time series is necessary to increase the forecasting quality. Second, such results may be caused by the fact that the training data set was very small (only 2222 negative cases and 66 positive cases used for training ( Table~\ref{table1}). On the other hand, our conclusions indicate a possibility to validate the forecast on longer time series using the proton flux and SXR data only, or to use the simplified characteristics of the ARs \citep[such as the Hale and Mackintosh classes,][]{McIntosh1990} instead of the vector magnetic field characteristics.
        
    Table~\ref{table3}  highlights how valuable various groups of characteristics are for prediction. The most critical set of characteristics comes from the characteristics of the preceding $>$\,10\,MeV proton flux. In fact, the forecast is not far, in terms of performance, from one involving all characteristics even if only SPE characteristics are used. The prediction becomes significantly worse if these characteristics are excluded from the forecast, while predictions based solely on these characteristics are relatively good. This is understandable as the prediction captures the persistence in SPE occurrence. Essentially, it may capture the fact that if the SPE is happening today and does not end by the end of today, it continues tomorrow. This is also evident from the experiment when we removed the last $>$10\,MeV proton flux value for the day (at 22:00\,UT). As Table~\ref{table3} indicates, the performance of proton flux-based forecast drops in this case.
        
    It was also evident that excluding any of the other groups of characteristics (AR, SXR, or radio burst) does not affect the prediction. We tested how the inclusion of any one among these three groups, in addition to the proton flux characteristics, affects the forecast. The results are shown in Figures~\ref{figure:ROCvar}\textit{c-e} and Table~\ref{table3}. The inclusion of either the SXR or AR group of characteristics in addition to the proton flux characteristics produces predictions with scores comparable with the forecast based on all characteristics. The inclusion of both these groups simultaneously does not improve the prediction. Also, the inclusion of the radio bursts together with the proton flux characteristics does not improve the prediction but even makes it worse. In some sense, the counts of the radio bursts seem to be the least valuable feature for the forecast within the settings of this work. Inclusion of the SXR and AR characteristics increases the prediction performance, with a higher preference for SXR characteristics. It is well known that the average SXR levels (and corresponding flare productiveness) are related to the complexity of ARs \citep{Falconer2012,Sadykov2017}. Therefore, such results are not surprising. The only significant difference between these characteristics is when a complex SPE-productive AR is behind the western limb, which was encountered rarely during the Solar Cycle 24 (\url{https://umbra.nascom.nasa.gov/SEP/}).
    
    Another interesting observation following from Table~\ref{table3} is that the inclusion of additional characteristics or increasing the complexity of the neural network does not necessarily lead to an increase of the forecast performance and may even result in a notable decrease. One can notice this by comparing 1) the network trained on the AR characteristics only and the network where SXR and radio bursts were introduced in addition (lines 3 and 8 in Table~\ref{table3}); 2) the network trained on a full set of characteristics and on SXR and proton flux characteristics only (lines 1 and 13 in Table~\ref{table3}); or 3) the AR-trained network where the AR Block outputs are summed and not summed before propagating into the main network (lines 3 and 6 in Table~\ref{table3}). This behavior may have two reasons behind it. First, the introduction of new characteristics should not necessarily improve the forecast if the characteristics are not so relevant to the forecast. The effects of the metrics decreasing with the increase of the number of characteristics are evident, for example, in Figure~4 by \citet{Bobra2015}. Second, the train data set have a very small number of positive samples. In this situation, the introduction of additional degrees of freedom into the forecasting model in terms of characteristics or network structure may result in a strengthening of memorization effects.
        
\begin{figure}[ht]
	\centering
	\includegraphics[width=1.0\linewidth]{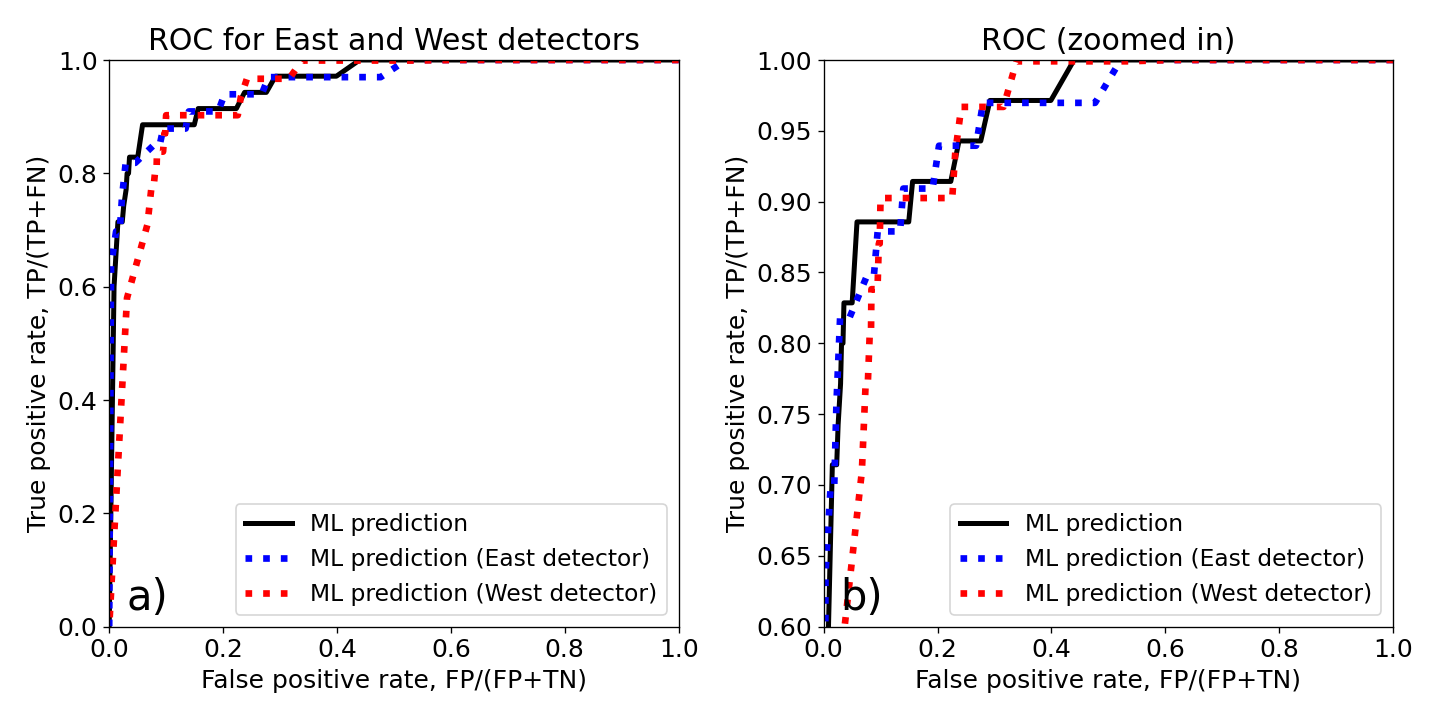}
	\caption{(a) Comparison of the ROC curves for the ML predictions using the averaged particle flux from the east and west GOES detectors (black solid line), using only the east detector particle flux (blue dotted lines), and using only the east detector particle flux (red dotted lines). Panel (b) presents the zoomed version of the same ROC curves.}
	\label{figure:EW}
\end{figure}

    While constructing SPE predictions in this work, we have used the flux of the $>$\,10\,MeV protons averaged over the east-oriented and west-oriented measurements of protons by the GOES satellite \citep{Rodriguez2010}. Because the distribution of particles is anisotropic, the measurements of the two directions are not exactly equal to each other. How does using the data from only one of the directions affect the forecasts? Figure~\ref{figure:EW} and Table~\ref{table4} show that using the particle flux data from one of the two detectors instead of averaging them changes the number of days with SPEs, and the overall performance of the machine learning approach slightly decreases. On the one hand, this demonstrates that the usage of the detector-averaged flux leads to a more robust prediction. On the other hand, it shows that using data from the GOES satellites that did not have measurements in two directions does not result in a significant loss of performance, especially in the regime of all-clear forecasts.
        
    Although we have formulated the prediction of SPEs as a simplistic feature-based binary classification problem in this study, it leads to several important outcomes and analysis insights. The Weighted True Skill Statistics (WTSS) score can tell the end-user more about the applicability of the constructed prediction for various purposes. Our results show that the machine learning forecast performs better than the SWPC NOAA forecast when the missed events are much less desirable than the false alarms (i.e., when parameter $\alpha$ $>$ 1 for WTSS($\alpha$), Figure~\ref{figure:WTSS}). The same conclusion follows from the upper part of the ROC curves in Figure~\ref{figure:ROCgen}. The forecasting regime when missing events is not desirable is critical from the operational perspective to approach ``all-clear'' forecasts. Sometimes, missing a single SPE may lead to a hazardous situation and significantly harm operational planning. We see that the ML-based predictions deal with such a situation better than the current daily operational forecasts.
        
    Unfortunately, the current SWPC NOAA forecasts do not capture some SPEs at all. In 14 out of 101 SPE days in our data set, only 1\% probability of the event for that day was issued. In addition, there was a zero chance to catch these events by adjusting the probability threshold. On the other hand, the ML-based predictions demonstrated a capability to capture all SPEs with about a 40\% false-positive rate (correspondingly, about 500 days in the test data set were predicted as having SPEs while they did not). The false-positive rate for all-clear forecast becomes lower than 30\% if constructed based on the proton and SXR flux properties only. Although these numbers are still large, we note that a good portion of the test data set contained the peak of Solar Cycle 24 (years 2014-2015), when the solar activity was high almost every day, and complex ARs existed at the solar surface. It also contained the year 2017, when major SPEs happened on the Sun (and were captured by the algorithm, as Figure~\ref{figure:timeroll} demonstrates). Correspondingly, all-clear forecasts based on machine learning are possible, although with about a 30-40\% false-positive rate. Nevertheless, this seems to be a reasonable price for the all-clear approach.

\section{Conclusion}
\label{Section:conclusion}

    In this study we have implemented a novel approach to predict SPEs using a neural network of custom architecture that efficiently tackles the problems of whole-Sun prediction and unknown host active regions. The predictions were made based on the daily characteristics of the vector magnetic fields in ARs, emitted soft X-ray and proton fluxes, as well as the counts of type II, III, and IV radio bursts. The primary results of the investigation are the following:
    
    \begin{itemize}
        \item For a feature-based binary prediction, it is very important to investigate how the prediction behaves with respect to the different metrics and check for the forecast ROC curves or similar types of characteristics;
        \item If the prediction is constructed based on the AR characteristics only, then the prediction significantly benefits from the inclusion of proxies characterizing ARs on the western limb and the far side;
        \item Characteristics of the preceding 10\,MeV proton flux are the most valuable characteristics for the prediction. The inclusion of the soft X-ray flux characteristics and the AR characteristics is valuable, with a preference of SXR characteristics, while the counts of type II, type III, and type IV radio bursts are the least-valuable characteristics;
        \item Exclusion of the information about the AR characteristics and the radio bursts does not affect the forecast and even makes it better. This provides a possibility to construct an operational forecast based on the GOES observations only. The possible reasons for such behavior are discussed in Section~\ref{Section:discussion};
        \item The constructed machine-learning-based forecast is very promising in situations when missing events is very undesirable. For any $\alpha{} > $1 for $WTSS(\alpha{})$ metrics, the ML-based forecast outperforms the SWPC NOAA forecast.
    \end{itemize}
    
    The presented study has certain limitations. One of the most important limitations is the significant data shortage. We have analyzed the SPEs and assessed the prediction algorithm performance only for Solar Cycle 24, which contained just 101 SPE days (positive samples) in total. Such a short data set represents a challenge for the learning algorithm. Testing of the algorithm on a more extensive data set should lead to more generalized and robust conclusions. The data shortage is also the reason why we have utilized the definitive SHARP CEA data series instead of their near-real-time analogs \citep{Hoeksema2014-NRT,Nishizuka2021-NRTforecast,Georgoulis2021-FLARECAST}, since the routine computations of NRT data started only in September 2012. Another limitation is the way the machine learning procedure was set up in this work. Typically, the data sets are divided into three parts (train, validation, and test). The ML algorithm is tuned on the validation data set first and evaluated on the test data set then. Here, because of the data shortage, we have separated the data set into train and test parts only. One can assume that we have emulated the situation when the validation data set is entirely identical to the test data set. Because the same assumption was effectively made for the SWPC NOAA forecasts (i.e., the probability thresholds were optimized on the test part but not on the train part), we can still compare the performances of both approaches and generalize the conclusions.
    
    We also would like to indicate numerous possibilities to improve the analysis performed in this work. First, the prediction algorithm constructed here is based on daily characteristics. However, the actual data (from which the characteristics were derived) constitute a multivariate time series (MVTS). The potential of MVTS analysis for the prediction of SPEs still has to be understood. Another way to improve the prediction algorithm is to include more valuable data sources. In particular, it was found in several studies that the characteristics of Coronal Mass Ejections (CMEs) are connected to the characteristics of SPEs \citep{Richardson2014,Wang2019,Kalaivani2020,Bruno2021}. Therefore, although the CME observations are not available in real-time, the inclusion of CME characteristics available for the previous day may improve SPE forecasts. Other data sources of interest are the characteristics of proton fluxes of energies other than $>$10\,MeV, soft X-ray characteristics of individual solar flares \citep[such as temperatures and emission measures, ][]{Sadykov2019,Kahler2020}, the timing of the recorded radio bursts, etc. To make the prediction algorithm more robust and generalize it for the other solar-cycle conditions, we plan to extend the prediction to the Solar Cycle 23 data and work with other characteristics of ARs \citep[Hale and Mackintosh classes,][]{McIntosh1990}. Finally, we plan to extend the proposed forecast implementation to other definitions of SPEs (other particle flux and energy thresholds) and other prediction time windows. To summarize, the prediction of SPEs targeted to a particular goal (e.g., an ``all-clear'' SPE prediction as in this work) is still a challenging problem from many perspectives, and we plan to continue and broaden this effort in future work.

\acknowledgments

    The research was supported by NASA Early Stage Innovation program grant 80NSSC20K0302, NASA LWS grant 80NSSC19K0068, NSF EarthCube grant 1639683, and NSF grant 1835958. VMS acknowledges the NSF FDSS grant 1936361 and NSF grant 1835958. EI acknowledges the RSF grant 20-72-00106.

\bibliographystyle{aasjournal}

\bibliography{SEPprediction}

\newpage

\newpage

\newpage

\newpage

\newpage

\newpage

\newpage       

\newpage   

\newpage
\begin{table}
    \centering
	\caption{Properties of the train and test data sets.}
	\label{table1}
	\begin{tabular}{|c|c|c|}
	    \hline
	    Property    &   Train data set  &   Test data set   \\
	    \hline
	    Time covered    &   Jun 2010 - Dec 2013,   &   Jan 2014 - Dec 2015,   \\
	                    &   Jan 2016 - Dec 2016,   &   Jan 2017 - Oct 2018   \\
	                    &   Oct 2018 - Dec 2019    &      \\
	   \hline
	    SPE days        &   66    & 35   \\
	    \hline
	    Non-SPE days    &   2222    &   1178    \\
	    \hline
	    Class ratio     &   about 1/34    &     about 1/34   \\
	    \hline
	\end{tabular}
\end{table}

\newpage
\begin{table}
	\centering
	\caption{ML-based prediction model performance, and its comparison with SWPC NOAA forecasts and a persistence model.}
	\label{table:comparisonSWPC}
	\begin{tabular}{|c|c|c|c|}
	    \hline
	    Score    &   ML-based model & SWPC NOAA forecast    &   Persistence model   \\
	    \hline
	    HSS1                &   0.377$\pm$0.028 &   0.571   &   0.314   \\
	    HSS2                &   0.635$\pm$0.008 &   0.748   &   0.647     \\
	    Area below ROC      &   0.959$\pm$0.001 &   0.919   &   -    \\
	    Cross-entropy loss  &   0.273$\pm$0.006 &   0.461   &   -    \\
	    TSS                 &   0.820$\pm$0.006 &   0.786   &   0.647    \\
	    WTSS(2)             &   0.803$\pm$0.005 &   0.758   &   0.536    \\
	    WTSS(5)             &   0.857$\pm$0.004 &   0.765   &   0.425    \\
	    WTSS(10)            &   0.920$\pm$0.004 &   0.817   &   0.375    \\
	    \hline
	\end{tabular}
\end{table}

\newpage
\begin{sidewaystable}
	\centering
	\caption{Results of the tests of inclusion/exclusion of various characteristics for ML-based prediction model.}
	\label{table3}
	\footnotesize
	\begin{tabular}{|c|c|c|c|c|c|c|c|c|}
	    \hline
	    Model    &   HSS1   &   HSS2 &   Area below ROC  &   Cross-entropy loss  &   TSS    &   WTSS(2) &   WTSS(5) &   WTSS(10) \\
	    \hline
	    ML model (baseline)             &   0.377$\pm$0.028 &   0.635$\pm$0.008 &   0.959$\pm$0.001 &   0.273$\pm$0.006 &   0.820$\pm$0.006 &   0.803$\pm$0.005   &   0.857$\pm$0.004   &   0.920$\pm$0.004   \\
	    SWPC NOAA forecast       &   0.571 &   0.748 &   0.919 &   0.461 &   0.786 &   0.758   &   0.765   &   0.817   \\
	    ML AR-based                  &   0.0             &   0.294$\pm$0.010 &   0.907$\pm$0.002 &   0.379$\pm$0.004 &   0.736$\pm$0.004 &   0.803$\pm$0.006   &   0.873$\pm$0.003   &   0.912$\pm$0.003   \\
        ML AR-based (no coordinates) &   0.0             &   0.331$\pm$0.014 &   0.915$\pm$0.002 &   0.395$\pm$0.008 &   0.723$\pm$0.010 &   0.795$\pm$0.008   &   0.874$\pm$0.004   &   0.927$\pm$0.007   \\
        ML AR-based (no extension)   &   0.0             &   0.067$\pm$0.033 &   0.655$\pm$0.026 &   0.642$\pm$0.004 &   0.333$\pm$0.008 &   0.536$\pm$0.006   &   0.755$\pm$0.008   &   0.866$\pm$0.004   \\
        ML AR-based (ARs distinguished)   &   0.0             &   0.110$\pm$0.014 &   0.798$\pm$0.018 &   0.520$\pm$0.016 &   0.559$\pm$0.036 &   0.676$\pm$0.016   &   0.834$\pm$0.011   &   0.910$\pm$0.006   \\
        ML (no AR data)              &   0.383$\pm$0.003 &   0.653$\pm$0.008 &   0.961$\pm$0.001 &   0.267$\pm$0.002 &   0.800$\pm$0.003 &   0.808$\pm$0.004   &   0.883$\pm$0.004   &   0.936$\pm$0.003   \\
        ML (no proton flux data)                &   0.011$\pm$0.013   &   0.244$\pm$0.024 &   0.854$\pm$0.004 &   0.494$\pm$0.012 &   0.595$\pm$0.014 &   0.655$\pm$0.015   &   0.800$\pm$0.005   &   0.888$\pm$0.001   \\
        ML (no SXR flux data)                &   0.406$\pm$0.033 &   0.634$\pm$0.021 &   0.954$\pm$0.001 &   0.277$\pm$0.007 &   0.814$\pm$0.013 &   0.810$\pm$0.013   &   0.849$\pm$0.011   &   0.909$\pm$0.014   \\
        ML (no radio burst data)        &   0.423$\pm$0.021 &   0.642$\pm$0.012 &   0.961$\pm$0.003 &   0.263$\pm$0.008 &   0.825$\pm$0.016 &   0.808$\pm$0.016   &   0.877$\pm$0.009   &   0.928$\pm$0.010   \\
        ML proton flux-based                    &   0.571$\pm$0.018 &   0.748$\pm$0.008 &   0.946$\pm$0.001 &   0.283$\pm$0.003 &   0.808$\pm$0.007 &   0.777$\pm$0.005   &   0.834$\pm$0.006   &   0.883$\pm$0.003   \\
        ML proton flux-based (no last value)    &   0.540$\pm$0.048 &   0.656$\pm$0.017 &   0.913$\pm$0.019 &   0.313$\pm$0.010 &   0.800$\pm$0.006 &   0.771$\pm$0.004   &   0.779$\pm$0.046   &   0.845$\pm$0.032   \\
        ML proton+SXR-based                &   0.434$\pm$0.046 &   0.673$\pm$0.015 &   0.965$\pm$0.001 &   0.256$\pm$0.002 &   0.821$\pm$0.003 &   0.830$\pm$0.003   &   0.915$\pm$0.002   &   0.954$\pm$0.001   \\
        ML proton+AR-based                &   0.440$\pm$0.039 &   0.650$\pm$0.012 &   0.949$\pm$0.003 &   0.286$\pm$0.011 &   0.813$\pm$0.010 &   0.783$\pm$0.006   &   0.850$\pm$0.016   &   0.915$\pm$0.015   \\
        ML proton+radio burst-based          &   0.423$\pm$0.028 &   0.664$\pm$0.019 &   0.940$\pm$0.001 &   0.303$\pm$0.003 &   0.806$\pm$0.011 &   0.794$\pm$0.007   &   0.835$\pm$0.002   &   0.858$\pm$0.001   \\
        \hline
    \end{tabular}
\end{sidewaystable}

\newpage
\begin{table}
	\centering
	\caption{Comparison of ML-based predictions trained on the particle flux data from east/west GOES detectors.}
	\label{table4}
	\begin{tabular}{|c|c|c|c|}
	    \hline
	    Property/Score    &   east/west averaged flux & east detector flux    &   west detector flux   \\
	    \hline
	    SPE days            &   101             &   109               &     100 \\
	    Non-SPE days        &   3400            &   3392              &     3401    \\
	    HSS1                &   0.377$\pm$0.028 &   0.498$\pm$0.024   &   0.096$\pm$0.122   \\
	    HSS2                &   0.635$\pm$0.008 &   0.709$\pm$0.014   &   0.354$\pm$0.047   \\
	    Area below ROC      &   0.959$\pm$0.001 &   0.958$\pm$0.001   &   0.941$\pm$0.003   \\
	    Cross-entropy loss  &   0.273$\pm$0.006 &   0.299$\pm$0.005   &   0.319$\pm$0.007   \\
	    TSS                 &   0.820$\pm$0.006 &   0.793$\pm$0.003   &   0.799$\pm$0.004   \\
	    WTSS(2)             &   0.803$\pm$0.005 &   0.795$\pm$0.005   &   0.803$\pm$0.001   \\
	    WTSS(5)             &   0.857$\pm$0.004 &   0.852$\pm$0.005   &   0.888$\pm$0.006   \\
	    WTSS(10)            &   0.920$\pm$0.004 &   0.913$\pm$0.005   &   0.938$\pm$0.003   \\
	    \hline
	\end{tabular}
\end{table}

\end{document}